# Measuring transient reaction rates from non-stationary catalysts

Dmitriy Borodin[1,2], Kai Golibrzuch[2], Michael Schwarzer[1], Jan Fingerhut[1], Georgios Skoulatakis[2], Dirk Schwarzer[2], Thomas Seelemann[3], Theofanis Kitsopoulos[1,2,4,5]* and Alec M. Wodtke[1,2,6]**

[1]Institute for Physical Chemistry, Georg-August University of Goettingen, Tammannstraße 6, 37077 Goettingen, Germany.

[2]Department of Dynamics at Surfaces, Max Planck Institute for Biophysical Chemistry, Am Fassberg 11, 37077 Goettingen, Germany.

[3]LaVision GmbH, Anna-Vandenhoeck-Ring 19, 37081 Goettingen, Germany.

[4]Department of Chemistry, University of Crete, Heraklion, Greece

[5]Institute of Electronic Structure and Laser – FORTH, Heraklion, Greece

[6]International Center for Advanced Studies of Energy Conversion, Georg-August University of Goettingen, Tammannstraße 6, 37077 Goettingen, Germany.

Email: *theo.kitsopoulos@mpibpc.mpg.de, **alec.wodtke@mpibpc.mpg.de

## Abstract

Up to now, the methods available for measuring the rate constants of reactions taking place on heterogeneous catalysts require that the catalyst be stable over long measurement times. But catalyst are often non-stationary, they may become activated under reaction conditions or become poisoned through use. It is therefore desirable to develop methods with high data acquisition rates for kinetics, so that transient rates can be measured on non-stationary catalysts. In this work, we present velocity resolved kinetics using high repetition rate pulsed laser ionization and high-speed ion imaging detection. The reaction is initiated by molecular beam pulses incident at the surface and the product formation rate is observed by a sequence of laser pulses at a high repetition rate. Ion imaging provides the desorbing product flux (reaction rate) as a function of reaction time for each laser pulse. We demonstrate the method using a 10 Hz pulsed CO molecular beam pulse train to initiate CO desorption from Pd(332)—desorbing CO is detected every millisecond by non-resonant multiphoton ionization using a 1−kHz Ti:Sapphire laser. This approach overcomes the time consuming scanning of the delay between CO and laser pulses needed in past experiments and delivers a data acquisition rate that is $10 − 1000$ times higher. We also apply this method to CO oxidation on Pd(332)—we record kinetic traces of $CO_2$ formation while a CO beam titrates oxygen atoms from an O-saturated surface. This provides the reaction rate as a function of O-coverage in a single experiment. We exploit this to produce controlled yet inhomogeneously mixed reactant samples for measurements of reaction rates under diffusion-controlled conditions.

## 1. Introduction

Methods to measure the kinetics of surface reactions are crucial to improving our understanding of heterogeneous catalysis. Traditionally, temperature programmed reaction, molecular beam relaxation spectrometry and phase-lag detection have been available to experimentalists [1-4]. Recently, the kinetic trace was obtained using velocity resolved methods [5] based on ion imaging [6-8]. This is essentially a pump-probe technique where a molecular beam pump-pulse initiates the reaction and pulsed laser ionization probes the desorbing products. Varying the delay between the two pulses provides the time base of the reaction kinetics. The ionized products are recorded with ion imaging providing product velocity information with every detection pulse. This allows measured product densities to be converted to product flux, which is by definition the reaction rate for a surface reaction. Furthermore, flight-times irrelevant to the reaction time can be subtracted from the experimental time-axis [9]. Like all pump-probe measurements, during the time that the delay between pump and probe is being scanned, the reacting system under study must not change. However, catalysts are often dynamic. Catalyst composition can change dramatically under reactive conditions[10]—living catalyst [11-13]—and catalytic use can lead to poisoning [14]. For such systems, we need methods that can rapidly obtain kinetic information while the catalyst is changing.

In this work, we demonstrate velocity resolved kinetics with high repetition rate detection. The reaction starts when a pulse of molecules arrives at the surface and ion images are recorded for each pulse of a high repetition-rate laser that ionizes desorbing products. The ion images preserve the velocity information from which the rate of reaction is derived. The inverse repetition rate of the laser sets the temporal resolution. We demonstrate a duty cycle that is one to three orders of magnitude higher than previous methods [5], allowing measurements on a changing catalyst. The present experiments use a 1-kHz Ti:Sapphire laser—future experiments with Yb-fiber lasers operating at $10^{2-3}$ kHz provide a perspective for improvement.

## 2. Experiment

We previously described the apparatus in detail elsewhere [6-8]. Briefly, we produce two molecular beams in two vacuum chambers, each equipped with piezo-electrically actuated pulsed valves. The valves' repetition rates are variable up to 500 Hz. The pulse durations can be as low as 30 μs. Each beam passes through two differential pumping chambers, before entering an ultrahigh vacuum (UHV) chamber with a base pressure of $2\times10^{-10}$ mbar, where they intersect one another and collide with a Pd(332) surface. One beam collides at normal incidence dosing the sample with oxygen. The second beam, incident at 30° to the normal, initiates the reaction with a pulse of CO. The CO beam can either be used alone to study CO trapping/desorption or with an oxidized surface to initiate $CO_2$ formation. A single crystal of Pd cut and polished to expose the (332) surface is mounted on a 5-axis manipulator and can be heated to 1150 K using electron bombardment.

The instrument is equipped with an Ar⁺ sputtering source for cleaning the surface as well as an Auger electron spectrometer to check its cleanliness.

A homogeneous electric field oriented parallel to the surface is formed by two parallel flat meshes (repeller and extractor), between which both molecular beams pass. After ionization of the reaction products by a non-resonant multiphoton process, using an ultra-short Ti:Sapphire laser (Coherent Astrella, 800 nm, 35 fs, 0.5 mJ, 1 kHz) focused with a 150 mm plano-convex lens, a 3 kV pulse applied to the repeller of the ion imaging system directs the ions to the imaging detector. This maps the products' density and in-plane velocity vectors, which is used to create a flux image. A region of the flux image is then integrated to provide the rate of reaction at a specific time. We record ion images with a 56 mm Chevron MCP detector coupled to a P43-phosphor screen, whose phosphorescence detected by a high-frame rate CMOS camera (Vision Research Phantom VEO 710). We took advantage of commercial data acquisition software (DaVis LaVision GmbH) and a software-controlled timing unit (PTUX, LaVision GmbH). The timing unit is triggered both at 10 Hz—synchronized with the pulsed nozzle—and at 1−kHz—synchronized with the laser. Several thousand images are recorded over several seconds and stored on the camera's internal memory, only to be transferred later to a computer's hard disk.

Figure. 1 shows a comparison to methods requiring the delay between pulsed molecular beam and laser, $t_{BL}$, to be scanned. In that case (Fig. 1a), an ion image is measured for a fixed value $t_{BL}$ and an ion image is accumulated over many (typically 50) molecular beam pulses. $t_{BL}$ is then incremented and the process is repeated. Here, one ion image is recorded for every molecular beam pulse, whose repetition rate is typically 10 to 100 Hz. Using high repetition-rate detection, the ion image is recorded every millisecond. Each pulse of the laser (points in Fig. 1b) corresponds to a point in the temporal evolution of the reaction. The P43 phosphor screen decays over $\tau_{90\% \to 10\%} = 1.3$ ms, while the time between laser pulses is only 1 ms. Hence, after downloading the image sequence to the computer we subtracted from each image the "afterglow background" remaining from the previous image.

### 3. Results and Discussion

#### 3.1. Proof of principle: application to CO desorption from Pd(332)

As a proof of principle, we performed measurements on CO trapping/desorption from Pd(332) between 583-623 K. Here, a clean Pd(332) crystal is exposed to a pulsed molecular beam of pure CO operating at 10 Hz. The surface temperature is controlled so that a 1−kHz detection rate is sufficiently rapid to follow the desorption kinetics, while also ensuring that all CO molecules desorb between molecular beam pulses. Following Ref.'s [6-7], we extract the kinetic trace by integrating flux images between 300 and 900 m/s and

±4° form the surface normal. This captures most of the desorbing molecules, while suppressing signal from directly scattered (higher velocity) and background (lower velocity) CO.

Figure 2 shows data from a typical 5-second experiment, requiring 1% the measurement time needed for delay-scanning. 50 kinetic traces result, one from each of 50 CO molecular beam pulses. The inset shows three kinetic traces in detail. We filter to the raw data (blue) with a periodic Savitzky-Golay filter [15] applied by first sorting the data according to $t_{BL}$ and then employing a moving linear fit to a single data point and 10 of its neighboring data points—all with the same $t_{BL}$. The value of the fitted line then replaces the data point and the process is repeated on the next data point. This leads to the filtered output (black). The CO desorption rate constant, $k_d$, is determined by fitting each pulsed decay with a function that convolves the incident CO beam's temporal profile with an exponential decay—red line in Fig. 2 inset decay [5]. In this way, we derive fifty independent values of $k_d$, from which we obtain an average value and a standard deviation.

Figure 3 shows $k_d$ values for CO on Pd(332) and Pd(111) using several different methods. The rate constants obtained from the data of Fig. 2 (×) are in good agreement with other methods. We note that the observed desorption rates depend little on the presence of atomic steps that are found in high concentration on the Pd(332) surface [7, 16]. Clearly, steps do not significantly stabilize CO on Pd, a conclusion that is consistent with reported isosteric heats of adsorption [17]. An Arrhenius fit to $k_d$ values using Pd(332) results yields $E_a = 1.58 \pm 0.02$ eV and $A = 10^{15.6 \pm 0.3}$ s$^{-1}$.

### 3.2. Duty-cycle analysis

We consider now the quantitative duty-cycle improvements possible with high rep-rate detection, within the specific context of desorption rates near zero-coverage. We first define a characteristic desorption time, $\tau$, which is the inverse of the desorption rate constant, $\tau = k_d^{-1}$. This imposes an upper limit of the molecular beam's repetition rate ($f_{MB}^{max}$) and therefore a minimum repeat-time, $t_{min} = 1/f_{MB}^{max}$, needed to maintain the low coverage condition. While there is some ambiguity involved, we set $t_{min} = 5\tau$, the time at which a 1$^{st}$-order decay has reached 0.7% of its initial value. Data obtained within $t_{min}$ are most important to the fitting—we label this data "relevant".

The number of relevant data obtained from each molecular beam pulse used in the high rep-rate approach, $n_{HRR}$ is given by:

$$n_{HRR} = t_{min} \times f_L \qquad (1)$$

where $f_L$ is the detection laser repetition rate and data acquisition rate, $\dot{n}_{HRR}$, is given by:

$$\dot{n}_{\text{HRR}} = t_{\min} \times f_{\text{L}} \times f_{\text{MB}} \tag{2}$$

where $f_{\text{MB}}$ is the repetition rate of the pulsed molecular beam.

The number of relevant data per molecular beam pulse in a conventional delay-scanning experiments, $n_{\text{DS}}$, is 1 and the data acquisition rate is then:

$$\dot{n}_{\text{DS}} = f_{\text{MB}}. \tag{3}$$

Taking the ratio of these two data acquisition rates, we find that the theoretical improvement in duty cycle is given by:

$$\frac{\dot{n}_{\text{HRR}}^{\max}}{\dot{n}_{\text{DS}}^{\max}} = t_{\min} \times f_{\text{L}} = f_{\text{L}} \big/ f_{\text{MB}}^{\max}. \tag{4}$$

The data acquisition rate using the delay-scanning approach is limited by $t_{\min}^{-1}$ and, of course, technical limitations to the rep-rate of pulse beams (in our experience ~500 Hz), whereas $f_{\text{L}}$ is the only limiting factor to the data acquisition rate for the high rep-rate method. We emphasize that $f_{\text{L}}$ can be improved dramatically. This work used a Ti:Sapphire laser, $f_{\text{L}} = 1$ kHz; newly available Yb-fiber lasers achieve repetition rates of $10^2 - 10^3$ kHz, while still providing pulse energies and peak intensities sufficient for non-resonant multiphoton ionization.

The analysis so far neglects the number of ions produced in each experiment, which is equally important as the rate of data acquisition. All velocity resolved kinetics signals are proportional to the rate of product formation [6]. Hence in the desorption experiments presented above, the number of ions detected per laser pulse is proportional to $1/\tau$. The dependence on $\tau$ reflects the temporal dilution seen for slow reactions. Each molecular beam pulse deposits the same number of CO molecules on the surface; so, the observed density is diluted greatly over time for slow reactions and less so for fast reactions. Taking this into account, we may define the "count acquisition rate" (CAR).

$$\text{CAR} \equiv \frac{\dot{n}}{\tau} = \dot{n} \times k_{\text{d}}. \tag{5}$$

$$\text{CAR}_{\text{DS}} \equiv \frac{f_{\text{MB}}}{\tau}. \tag{5a}$$

$$\text{CAR}_{\text{HRR}} \equiv \frac{5\,\tau \times f_{\text{L}} \times f_{\text{MB}}}{\tau} = 5 \times f_{\text{MB}} \times f_{\text{L}}. \tag{5b}$$

This quantity determines the signal-to-noise (S/N) of the data obtained in any experiment. These equations point out that experiments using delay-scans exhibit decreasing S/N as $\tau$ increases, but this is not the case for high rep-rate measurements. This of course, mirrors the implications of Eq. 4. This also means that comparing different data acquisition methods should be done as a function of $\tau$.

Figure 4 shows calculated values of CAR vs. $\tau$ for a few different experimental configurations. Here, we only consider $\tau$ values larger than the shortest molecular beam pulse, which defines the kinetic resolution (black vertical line). To ensure that the CAR results only from relevant data, the molecular beam repetition rate should be matched to $t_{\min} = 5\tau = 1/f_{MB}$. This is true for either delay-scanning or high rep-rate detection. This gives rise to CAR plots for optimized delay-scanning (blue dashed line) and optimized 1−kHz detection (blue solid line) in Fig. 4. The red solid line shows CAR when using optimized 100−kHz detection. We also show in Fig. 4 the CAR vs. $\tau$ for un-optimized experiments. Specifically, we show CAR plot for a delay-scan experiment with a fixed 20 Hz molecular beam (green dashed line) as well as a 1−kHz detection experiment with a 10 Hz rep-rate molecular beam (magenta solid line).

The range of rates that can be measured with high S/N is much larger for high rep-rate detection than for delay scanning. Notice that for the optimized experiments, CAR is decreasing with $\tau$; thus, long lifetimes are harder to measure with high S/N than are short lifetimes. However, for delay scanning the CAR is proportional to $\tau^{-2}$ while for high rep-rate experiments it is proportional to $\tau^{-1}$. This is reflected in Fig. 4 through the slope of CAR vs. $\tau$ for delay-scan measurements, which is steeper than that of high rep-rate experiments. Furthermore, increasing $f_L$ further increases CAR. This shows that the high rep-rate method becomes extremely attractive for measuring slow rates. From our experience, the feasibility limit in an optimized delay scanning experiment is reached for $\tau \sim$10-40 ms. Delay scanning measurements under these conditions take on the order of 1h. The same limit is reached in a 1−kHz measurement when $\tau \sim$5 s, which can be extended to 500 s with 100−kHz detection. This shows that the high rep-rate detection approach can be applied to measure $\tau^{-1}$ values over ~7 orders of magnitude, whereas, delay scanning is limited to at most 3 orders of magnitude. High rep-rate detection thus enables measurements over a wider temperature range, providing more accurate Arrhenius parameters and greater sensitivity to non-Arrhenius behavior.

We also compare experimentally observed CARs obtained from our actual CO desorption experiments. The vertical gray dash-dotted line of Fig. 4 (marked with $\tau$ @ 593 K) represents the temperature at which the CO desorption experiments presented in Fig. 2 were carried out. Here, delay-scanning required 20 min to obtain ~250 relevant data, while 1−kHz detection provided ~70 relevant data in 10s. The derived rate constants were of similar accuracy for both methods. Normalizing to the number of relevant data points obtained, we find that CAR increased by a factor of ~30 for 1−kHz detection compared to delay scanning. Seen at the $\tau$-value at 593 K, the theoretical CAR plots (magenta line) and (green dashed line) show a theoretical enhancement factor that is close the observed enhancement.

### 3.3. A real time titration experiment for CO oxidation at Pd(332)

The velocity resolved kinetics experiment carried out with delay scanning provides time-resolved information by recording a signal arising from two pulses with a variable delay. Such experiments require that the system under study does not change between each pulse-pair; however, this requirement is often not fulfilled in surface chemistry. For example, catalysts can become poisoned with use by buildup of carbon [14] or other trace impurities. Furthermore, the composition of the surface can change under reactive conditions [10]. This also has an important implication for molecular beam experiments. For example, if we begin with the Pd(332) crystal used above for CO desorption, clean it and start dosing with the CO and $O_2$ pulses, the concentration of adsorbed oxygen, $[O^*]$, will change with time in a way that is determined by the competitive kinetics of $O_2$ dissociation and adsorption, CO adsorption, reaction and desorption. Thus, $[O^*]$ is a complex function of the two beam fluxes and the rates of each elementary process.

This has been shown in detail for CO oxidation on Pt(111) using delay scanning [6], where a steady state oxygen concentration, $[O^*]_{SS}$, is established over a period of few seconds. Velocity resolved kinetics exploit such steady-state conditions to investigate the reaction rate's dependence on oxygen coverage. Specifically, CO and $O_2$ pulsed beams run asynchronously at a controlled repetition rate ratio (RRR) to fix $[O^*]_{SS}$. Each new value of RRR gives a new value of $[O^*]_{SS}$ that is determined by a titration. The titration involves first saturating the Pt(111) surface with oxygen by running the $O_2$ beam for several minutes. This is known to produce an $O^*$ coverage of $[O^*]_{sat} = 0.25$ Monolayer (ML). We then turn off the $O_2$ beam and run many CO molecular beam pulses while monitoring the $CO_2$ formation rate at a specific time within the kinetic trace, $t_{BL}$, the Beam-Laser delay time. The $CO_2$ formation rate at the chosen $t_{BL}$ changes as more CO pulses react at the surface, eventually going to zero when all the $O^*$ is removed from the surface. However, from the titration measurement at a single $t_{BL}$ it is not possible to determine the oxygen coverage. This is because the transient rate of $CO_2$ formation becomes slower as $O^*$ is removed from the surface. By choosing only one specific $t_{BL}$ we miss the change of the kinetic trace as a function of titration time. To account for the change of the kinetic trace during the titration, measurements are repeated for various $t_{BL}$, and the titration curves are integrated over $t_{BL}$. The integral of such titration curves is proportional to the total oxygen coverage on the surface. We compare the integral from oxygen saturated surfaces with those obtained from a steady-state oxygen covered surface to determine the fraction of the total oxygen coverage that remains under steady state conditions. Clearly, this procedure is not optimal; ideally, one would like to know the kinetic trace at each point in the titration. While this is tremendously tedious and time consuming to perform with delay scanning, it is easily achieved with high rep-rate detection.

Figure 5 shows such a measurement carried out on Pd(332) at $T_S = 503$ K. Here, we first saturated the surface with oxygen by dosing with a 500 Hz $O_2$ molecular beam pulse for 5 minutes (total exposure $300 \pm 80$ ML). The high rep-rate raw data (blue lines) results from a CO pulsed beam operating at 50 Hz. With each CO titrant pulse, a certain amount of oxygen is removed from the surface, so that each kinetic trace probes a different O-atom surface coverage. Using a 51 point periodic Savitzky-Golay filter as described above we filtered the raw data (blue lines of Fig. 5) to yield the filtered data (black lines of Fig. 5). Two insets in Fig. 5 show representative kinetic traces at early and late times in the titration. We find that the signal amplitude decreases and the rate slows with increasing titration time, reflecting the consumption of oxygen with each subsequent CO pulse. The filtered data can be represented by a first order decay for the entire 30 s titration time and for temperatures between 473 and 533 K.

### 3.4. Reaction rate analysis at high oxygen coverages

This approach provides new information about the nature of the kinetics and improves the performance of the velocity resolved kinetics methods. One advantage is the ability to obtain rates of reactions at saturated oxygen coverage, where the absolute oxygen coverage is unambiguously defined. To demonstrate this, we apply a simple model previously suggested by Engel and Ertl to describe CO oxidation kinetics on Pd [18]. The model incorporates four processes:

$$F_t(O_{2,g}) \xrightarrow{S_{O_2}} 2\, O^* \quad\quad\quad (R1)$$

$$F_t(CO_g) \xrightarrow{S_{CO}} CO^* \quad\quad\quad (R2)$$

$$CO^* \xrightarrow{k_d} CO_g + * \quad\quad\quad (R3)$$

$$CO^* + O^* \xrightarrow{k_r} CO_{2,g} + 2* \quad\quad\quad (R4)$$

where X* indicates an adsorbed species X to the surface, * indicates a free adsorption site and $F_t$ is the time dependent flux provided by the molecular beams to the surface. $S_X$ is the sticking coefficient of the species X. Under conditions of excess oxygen, the effective 1$^{st}$-order rate constant, $k_{eff}$, is given by:

$$k_{eff} = k_r[O^*] + k_d \quad\quad\quad (6)$$

and the $CO_2$ formation rate is given by:

$$\frac{d[CO_{2,g}]}{dt} = k_r[O^*][CO^*]. \quad\quad\quad (7)$$

We take advantage of the fact that the velocity resolved kinetics signal is directly proportional to the rate of $CO_{2,g}$ formation and that during the first few CO pulses arriving at the surface, we probe a well-defined oxygen coverage $[O^*]_{sat} = 0.292$ ML [6]. We present examples of such kinetic traces in Fig. 6, averaged over the first 20 CO pulses for three values of $T_S$. We fit each trace to obtain a first order time constant ($\tau_{eff} = k_{eff}^{-1}$). Since $k_d$ is known—in fact, under these conditions desorption is not competitive and $k_{eff} = k_r[O^*]_{sat}$; see blue shaded area of Fig. 3—and coverage $[O^*] \equiv [O^*]_{sat}$, we can easily determine $k_r$ at all three values of $T_S$. The derived effective reaction rate constants ($k_r[O^*]_{sat}$) are shown in Fig. 7 as an Arrhenius plot. Note that in Fig. 6 for $T_S = 533$ K, the temporal resolution is insufficient to provide a reliable fit. This problem could be solved by repetitive measurements of this type at a variety of delays of the CO molecular beam pulse – says interleaved in 0.1 ms steps. One could also use a higher repetition rate laser. Alternatively, $k_r$ can be obtained simply from the amplitude of the kinetic trace, shown as arrows in Fig. 6, which is proportional to the initial rate of product formation. This allows the rate constants at all three temperatures to be placed on the same scale using the Arrhenius law. In Fig. 7, the values derived from initial rates (×) are placed on an absolute scale by comparison to the rate constants obtained by exponential fitting at the lower two temperatures (o). The best fit Arrhenius parameter for the reaction rate constant $k_r$ are $E_a = 0.76 \pm 0.02$ eV and $A = 10^{11.0 \pm 0.4}$ s$^{-1}$ ML$^{-1}$. These results are consistent with independently obtained results using delay scanning.

### 3.5. Diffusion limited surface reaction rates

In the course of studying the behavior high rep-rate detection, we made what were, at first, surprising observations. We found that near the end of titrations when the rate of $CO_2$ formation had nearly vanished, oxygen coverage remained on the surface in regions outside the crossing region of the two molecular beams. By translating the crystal in a direction perpendicular to the surface normal, we could observe a sudden increase of the $CO_2$ production rate. These observations indicated that a successful modelling of these experiments would require characterizing both the spatial as well as the temporal evolution of the reactant coverages.

In this section, we describe such modelling show that titration experiments often produce conditions where CO diffusion effects the rates of reactions.

We first imagine dividing the reacting surface into $j$ spatial elements. The concentrations of $CO^*$ and $O^*$ in spatial element $j$, defined as $[O^*]_j$ and $[CO^*]_j$, are given by reactive terms:

$$\left(\frac{d[CO^*]}{dt}\right)_j^{rct} = S_{CO,j} F_{t,j}(CO_g) - k_d[CO^*]_j - k_r[O^*]_j[CO^*]_j, \tag{8}$$

$$\left(\frac{d[O^*]}{dt}\right)_j^{rct} = 2 S_{O_2,j} F_{t,j}(O_{2,g}) - k_r[O^*]_j[CO^*]_j \tag{9}$$

and diffusive terms:

$$\left(\frac{d[CO^*]}{dt}\right)_j^{dif} = D_{CO,j}\Delta[CO^*]_j, \tag{10}$$

$$\left(\frac{d[O^*]}{dt}\right)_j^{dif} = D_{O,j}\Delta[O^*]_j. \tag{11}$$

The total rate is the sum of reactive (Eq. 8 and 9) and diffusive (Eq. 10 and 11) contributions. In Eq. 10 and 11, $D_n$ and $\Delta[n^*]_j$ are the species specific and concentration independent diffusion coefficients and concentration gradients, respectively, used in application of Fick's second law of diffusion. $F_{t,j}$ is the time dependent incoming flux to the spatial element $j$, produced by the molecular beams.

The dosing function $F_{t,j}$ is described with a periodic function (in time) that reassembles the spatial, $g(r_j)$, and temporal shape, $f(t)$, of our molecular beams. Specifically, we modelled it with:

$$f(t) = \cos^{2n}(\pi\, RR\, (t - t_0)), \tag{12}$$

where $RR$ is the repetition rate of the nozzle, $t_0$ the reference timing and $n$ is an integer chosen to best represent the temporal shape of the beam. Using ion imaging, we experimentally determined the spatial intensity profile of each molecular beam, from which we deduced their radial profiles. Both molecular beams have a nominal projected diameter of 2 mm. For $F_{t,j}$ we use a flattop Gaussian that resembles the experimentally determined radial profile, $g(r_j)$, of the beam, which is given by:

$$g(r_j) = \exp\left(-\left[\frac{r_j^2}{2\sigma^2}\right]^m\right), \tag{13}$$

where $m$ and $\sigma$ are parameters representing the shape of the experimental beam profile. The combined and normalized dosing function is then given by:

$$F_{t,j} = \frac{g(r_j)f(t)}{N}, \tag{14}$$

where $N$ is the normalization to define the observed molecular beam flux.

We made sure that both molecular beams overlap on the surface and checked this by ensuring that oxygen coverage remained symmetrically distributed around the molecular beams crossing point at the end of the titration measurement. Hence, we conclude that our experiments approximately preserve radial symmetry, which allows us to solve the diffusion equations, Eq.'s 10 and 11, in polar coordinates. The diffusion formalism is derived in the Appendix to this paper. The rate equations including diffusion and reaction are solved numerically using LSODA from the Fortran ODEPACK library[19]. The concentrations of CO* and O* in each spatial element $j$ are propagated in time.

To simulate measurements like those of Fig. 5, we initiate the model calculations with adsorbed oxygen produced by many pulses of the $O_2$ beam. This requires an initial $O^*$ spatial profile (black line of Fig. 8) that is much broader than the nominal $O_2$ beam profile (thick gray line of Fig. 8), as $O^*$ coverage quickly saturates near the center of the beam and after that only the wings of the $O_2$ beam add additional $O^*$. We simulated the spatial evolution of concentrations within a radial extent of 3 mm and with each spatial element, $j$, being 5 µm in size. The corresponding total $CO_2$ formation rate is given by summing the rate of each spatial element $j$ and weighting it by the respective area $A_j$, in the following manner:

$$\frac{d[CO_2]_t}{dt} = k_r \sum_{j=0}^{j_{max}} [O^*]_{t,j}[CO^*]_{t,j} A_j, \qquad (15)$$

where the area of $j^{th}$ spatial element is given by:

$$A_j = \pi(2j+1)r_0^2. \qquad (16)$$

The simulation accounts for the influence of reactions R - R as well as CO desorption and diffusion. The reaction rate constants were determined previously (see Sec. 3.1 and 3.3). Oxygen desorption is unimportant at these surface temperatures [20-21]. Oxygen diffusion is found to be unimportant under our conditions [22]. We estimated the diffusion coefficient for CO using an activation energy of 0.12eV from Ref. [23] and the fitted pre-factor for CO diffusion needed to obtain agreement with our measurements. The optimized pre-factor was $10^{-3.7\pm0.3}$ cm$^2$ s$^{-1}$. The CO diffusion rates we obtain in this way are consistent with previous measurements on Pd(111) – see the Appendix.

We derived the absolute incident beam fluxes from measurements of the steady state pressures of CO and $O_2$ in the UHV chamber combined with a knowledge of the chamber pumping speed. The model results were insensitive to the $O_2$ flux, but highly sensitive to the assumed CO flux. We found best agreement with experiment when using a CO flux ~30% smaller than that derived from our experimental estimate.

We used a 2nd order Langmuir expression for the coverage dependent sticking coefficient of $O_2$,

$$S_{O_2}([O^*]) = S_{O_2,0}\left(1 - \frac{[O^*]}{[O^*]_{max}}\right)^2 \qquad (17)$$

with $S_{O_2,0} = 0.4$ [16]. Best agreement with the experiment is achieved when an oxygen coverage independent sticking coefficient of $0.6 \pm 0.1$ is used for CO. We assume that the sticking probability of CO decreases linearly with CO coverage.

Fig. 9 (panel A) shows a comparison of this model to the titration experiment of Fig. 5. Note that the amplitude quickly decays over the first 300-400 CO beam pulses, thereafter decaying more slowly, behavior that is captured in the kinetic model. The transition between the fast and slow decay regions is accompanied by an increase of the baseline (shown in magnification in panel B of Fig. 9). This indicates a continuous production of $CO_2$. The experimentally observed increase of the baseline is also present in the model. Looking in more detail (panels C of Fig. 9), we find that the single pulse transient rate is decreasing with increasing titration time; furthermore, the transient rates are well reproduced by the kinetic model as is the continuous production of $CO_2$ seen in the later stages of the titration.

The qualitative behavior can be understood by recalling that the amplitude of the titration curve reflects the initial rate of $CO_2$ production, which is directly proportional to the oxygen coverage. With increasing titration time, the initial rate decreases indicating that the oxygen coverage is dropping. As a consequence of the reduced reaction rate at lower oxygen coverage, the lifetime of CO molecules on the surface increases, while the rate of CO adsorption remains constant. Since CO's desorption rate is slow at these temperatures, CO begins to build up from one molecular beam pulse to the next; this leads to quasi-continuous $CO_2$ formation and to baseline increase at later times in the titration.

We also performed a sensitivity analysis of the fit to the titration kinetics. The degree of rate control [24] exhibited by the kinetic parameter, $k_i$, is given by a sensitivity coefficient, $X_{i,t}$:

$$X_{i,t} = \frac{k_i}{R_t}\left(\frac{dR_t}{dk_i}\right)_{k_{j\neq i}}, \qquad (18)$$

where $R_t$ is the $CO_2$ formation rate. A high absolute value of $X_{i,t}$ indicates importance of the process to the reaction rate. A positive (negative) value of $X_{i,t}$ means that an increase of the rate parameter produces an increase (decrease) of the $CO_2$ formation rate. In Fig. 9 (panels D) we plot $X_{i,t}$ for reaction (purple, dotted), CO desorption (blue, dashed) and CO diffusion (green, dash-dotted).

The reaction between CO* and O* dominates the rate of product formation up to a titration time of about 7 sec, thereafter CO desorption and diffusion become increasingly important. Between 12 and 24 seconds, where the three processes are of similar importance, their influence appears at different points in the kinetic trace. Consider the kinetic traces found at ~23 seconds. Here, the beginning of the kinetic trace is dominated by the influence of the reaction, whereas diffusion and desorption influence later times in the trace. Note that desorption decreases while diffusion *increases* the rate of $CO_2$ production. This can be understood by realizing that at later stages of the titration, O* has been depleted near the center of the CO beam. Each new CO pulse produces higher CO* concentration in the "doughnut hole" of O* concentration (see Fig. 8). These are the conditions where the quasi-continuous $CO_2$ formation rate (i.e. the $CO_2$ being produced prior to the next pulse) can appear as it is due to a diffusion-controlled reaction between CO* and O*.

Fig. 8 shows the model's predictions of the oxygen's spatial distribution at various times during the titration. As explained above, the initial oxygen coverage distribution (black solid line) on the surface is broader than the CO or $O_2$ beam's spatial profiles (gray solid line) used to dose the surface. At the early stages of titration (4 s — dotted black line), CO flux is highest near the beam center line where O* removal proceeds most rapidly. At 8 s (dashed black line) oxygen is removed near the center of the CO beam. In the central region of the spatial distribution where O* has now been depleted, CO's lifetime increases and begins building up from pulse to pulse. Hence, a spatially inhomogeneous "doughnut" reaction is produced with high CO coverage near the center of the beam and high O-coverage near the wings of the beam (see Appendix). The quasi-continuous $CO_2$ formation is produced at the intersection of the CO* and O* rich regions, forming a reaction front. The product formation rate at the reaction front depends not only on the reaction rate constant, but also on the CO diffusion coefficient. While the stationary $CO_2$ formation is from a diffusion-controlled reaction, the transient rate induced by a CO pulse at late titration times is due to direct population of oxygen rich regions from the outer flanks of the CO beam and is only slightly influenced by the mobility of the reactants.

With our validated kinetic model we can also estimate the associated reaction front speed (see Appendix) which is a characteristic property that can be measured for spatio-temporal pattern formation. In Fig. 10 the reaction front speed at 503 K is shown as a function of titration time. Prior to 6 seconds after the beginning of the titration, no reaction front is formed. However, from 6 to 8 seconds titration time a reaction front forms and its speed accelerated to 175 µm/s. With increasing titration time, the front speed decreases reaching speeds of around 10 µm/s at titration times longer than 15 s. We emphasize that the derived values of front speed are similar to those obtained in previous work for CO oxidation on Pt(110), which ranged from 1 to 100 µm/sec [25-26]. The fact that we derive front speeds nearly a factor of two higher than that work, probably results from faster thermal diffusion for CO on Pd compared to Pt [23, 27].

It is important to highlight that we have modeled the real time titration experiment without coverage dependent rate constants. Since we achieve good correspondence with the experiment, we claim that the rate constants have weak dependence on oxygen coverage in CO oxidation on Pd(332). However, this is in contradiction to the findings that were previously made on Pd(111) by Engel and Ertl [18]. We also find that our reaction rate constant is about 4-8 times higher than those reported from Pd(111). We think that steps lead to a higher reaction rate, consistent with previous observations on Pd [28] and Pt [6]. The reason why we have not taken reaction at steps and terraces explicitly into account is that we have not needed it for good match with the experiment. This is probably due to a rather fast exchange of CO and O atoms between terraces and steps which leads to an effective reaction rate composed of both reactions at steps and terraces.

We plan to investigate the details of the kinetic mechanism of CO oxidation at steps and terraces of Pd further in future.

## 4. Conclusions

This work shows how high-repetition rate lasers and ion imaging detection can be used to obtain the kinetic traces of catalytic processes from a single molecular beam pulse, overcoming the need for delay scans that are typical for pump-probe methods. The new approach provides an increased duty cycle resulting in rates of acquisition for kinetic data that are 10-1000 times faster than conventional delay scanning methods. The new method can measure rates over 5-7 orders of magnitude, dramatically better than when using delay scanning. The method is particularly attractive for measuring slow processes where temporal dilution would make delay scanning impossible.

This new approach can also be used to study catalytic reaction rates under conditions where the catalysts composition is changing under reactive conditions. We demonstrated this with a real time titration experiment, where the transient $CO_2$ production rates were obtained at many points in a CO oxidation titration experiment. Specifically, we showed that the transient rate of $CO_2$ formation could be measured from each subsequent CO pulse, where each pulse probes a different O-atom coverage on the surface. From accurate modelling of the titration experiment we are able to derive various rate constants relevant to CO oxidation on Pd(332). Of particular novelty, we easily found conditions where the CO oxidation was diffusion-controlled. The rate constants we found are summarized here:

$$\frac{k_r(T)}{\text{ML}^{-1}\,\text{s}^{-1}} = 10^{11.0 \pm 0.4} \exp\left(-\frac{0.76 \pm 0.02\text{ eV}}{k_B T}\right),$$

$$\frac{k_d(T)}{\text{s}^{-1}} = 10^{15.6 \pm 0.3} \exp\left(-\frac{1.58 \pm 0.02\text{ eV}}{k_B T}\right),$$

$$\frac{D_{CO}(T)}{\text{cm}^2\,\text{s}^{-1}} = 10^{-3.7 \pm 0.3} \exp\left(-\frac{0.12\text{ eV}}{k_B T}\right).$$

Our results are also consistent with an oxygen coverage independent sticking coefficient of CO of $0.6 \pm 0.1$. The desorption and diffusion rate constants of CO agree well with the parameters determined earlier from Pd(111), indicating that CO has no energetic preference for steps and that they are not influencing its mobility on the surface. The reaction rate constant is found to be approximately a factor 4-8 higher than previous reports for Pd(111), indicating that steps are more reactive for CO oxidation on Pd than terraces.

While in this work we were limited to a detection rep-rate of 1−kHz due to the fact that we used a Ti:Sapphire laser, we plan to extend our capabilities to a detection rate of 100 kHz and study reaction rates at changing catalysts conditions in more detail using a Yb-Fiber laser. We think that this method offers the possibility to accurately study catalytic reaction rates and kinetic mechanisms at the intersection between the well-defined conditions that are desirable for surface science and the more dynamic conditions relevant to industrial catalysis.

# 5. Figures

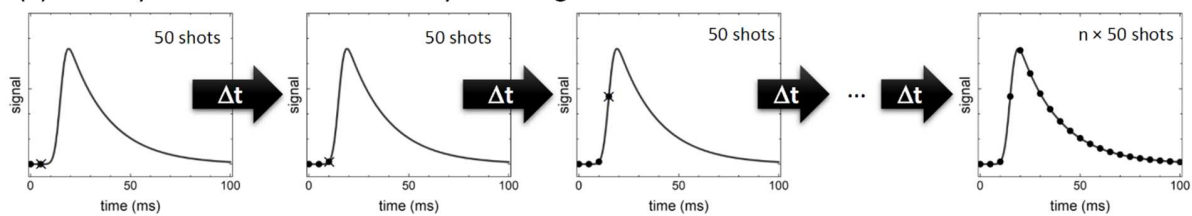

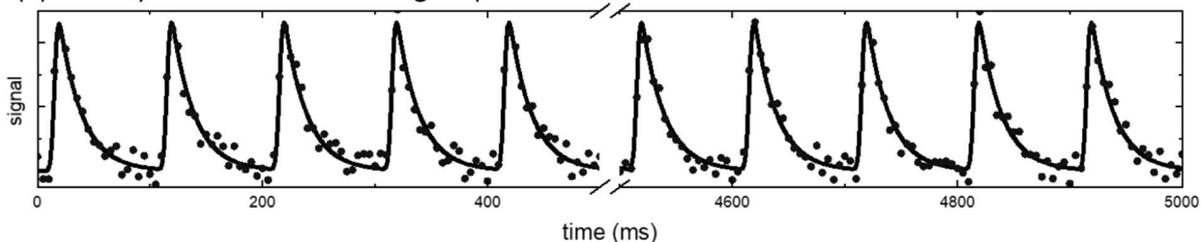

Fig. 1: Comparison of delay scanning versus high rep-rate detection employed in velocity resolved kinetics measurements. (a) Delay scanning involves the acquisition of many (e.g. 50) images at each time delay between the initiating molecular beam pulse and the laser ionization pulse. Points in the kinetic trace recorded by scanning the delay between a molecular beam pulse that initiates the reaction and a laser ionization pulse that detects the products. The catalytic system must be stable throughout the course of the delay scanning procedure. (b) High rep-rate detection with high-speed imaging records many points in the kinetic trace for each molecular beam pulse. Here, the molecular beam initates the reaction every 0.1 s and points in the kinetic trace are recorded by each pulse of a 1−kHz detection laser. The duty cycle of this method can be much higher than delay scanning. Furthermore, the kinetics can be recorded while the catalyst is changing.

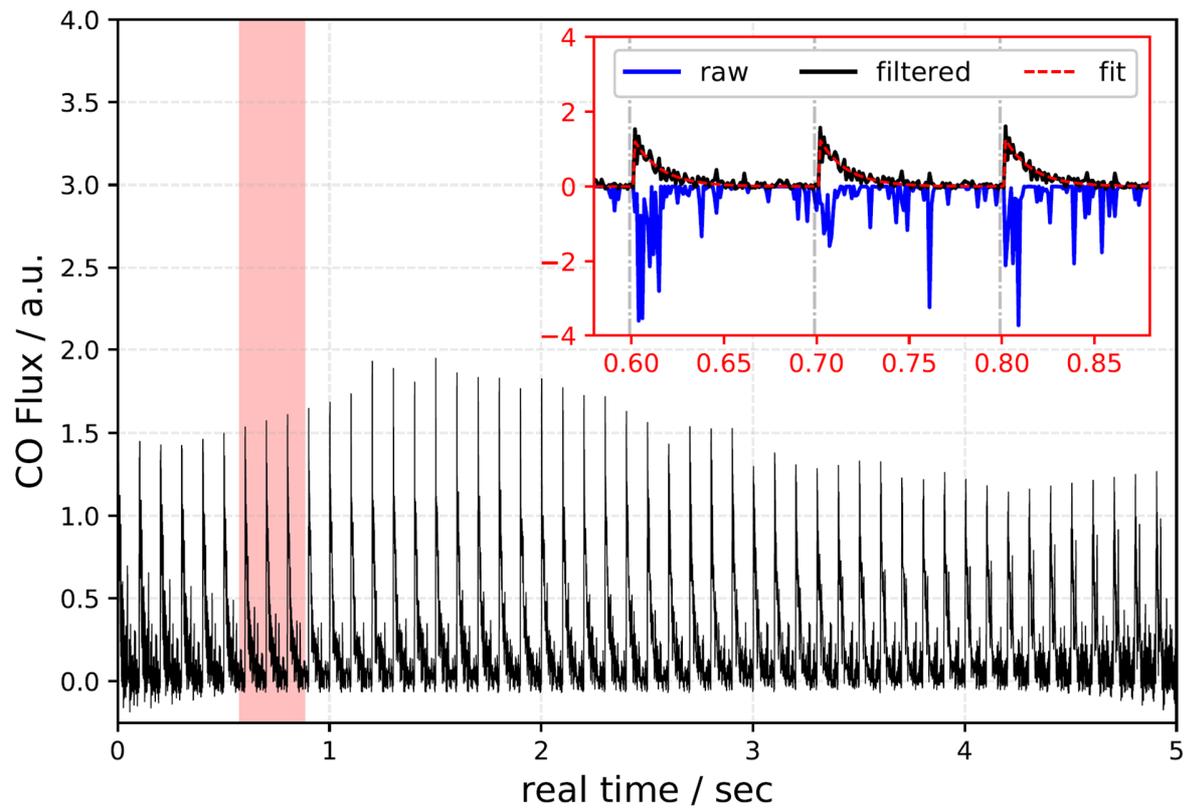

Fig 2: CO trapping/desorption from Pd (332) as a proof-of-principle for high rep-rate detection. The CO pulsed beam runs at 10 Hz, while the detection laser runs at 1−kHz. The Pd crystal was held at 593 K. Inset: Raw data (blue line) is treated by the Savitzky-Golay filter (see text) to yield the filtered data (black line). The grey dash-dotted line indicates the time at which the CO pulse initiates the reaction. The dashed red line is a periodic first order decay convoluted with the temporal profile of the CO beam and is used to extract the desorption rate constant.

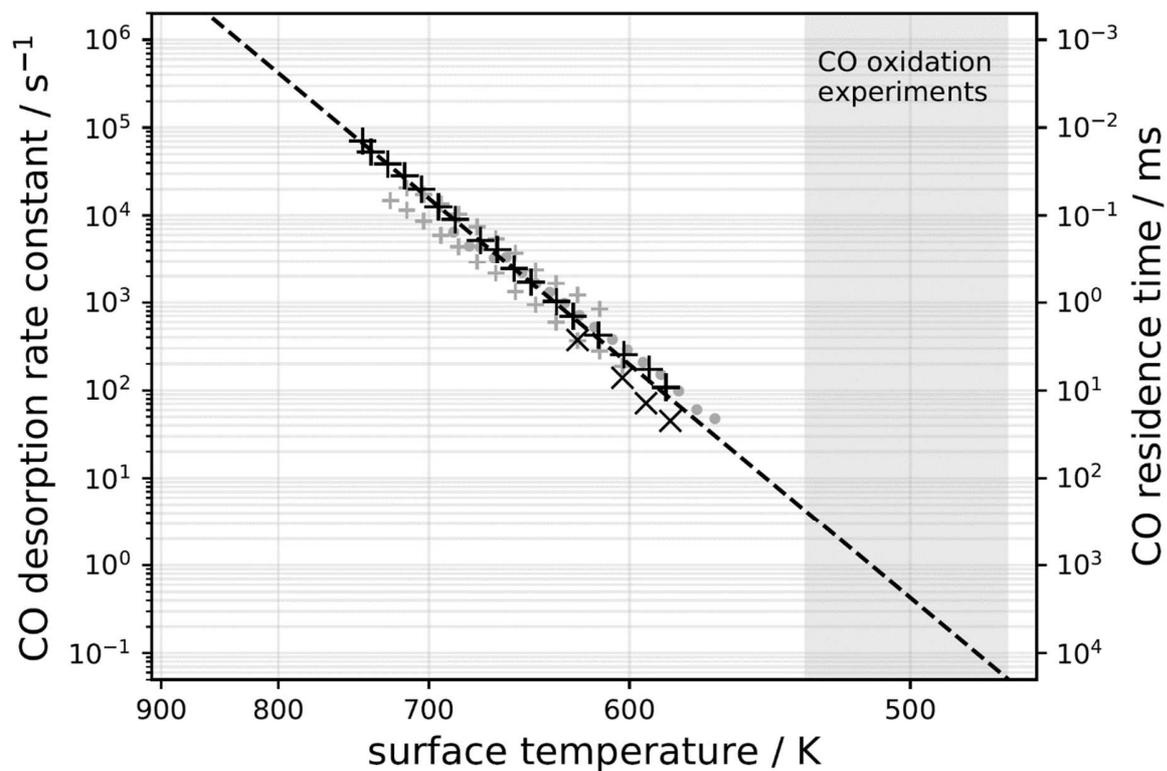

Fig 3: Desorption rate constants of CO from Pd(332) and Pd(111) vs. surface temperature. × indicates values obtained from high rep-rate detection . + indicates results from delay scanning. + used delay scanning with Pd(111) [7] and ● indicate results on Pd (111) from Modulated Molecular Beam Spectrometry [16]. The black dashed line is an Arrhenius fit ($A = 10^{15.6 \pm 0.3}\ s^{-1}$ and $E_a = 1.58 \pm 0.02\ eV$) to all desorption rate constants on Pd(332). The gray shaded region indicates the temperature range at which CO oxidation measurements in this work are conducted—see section 3.3. Uncertainties in the rate constants determined from delay scanning and high rep-rate detection are smaller than the symbols.

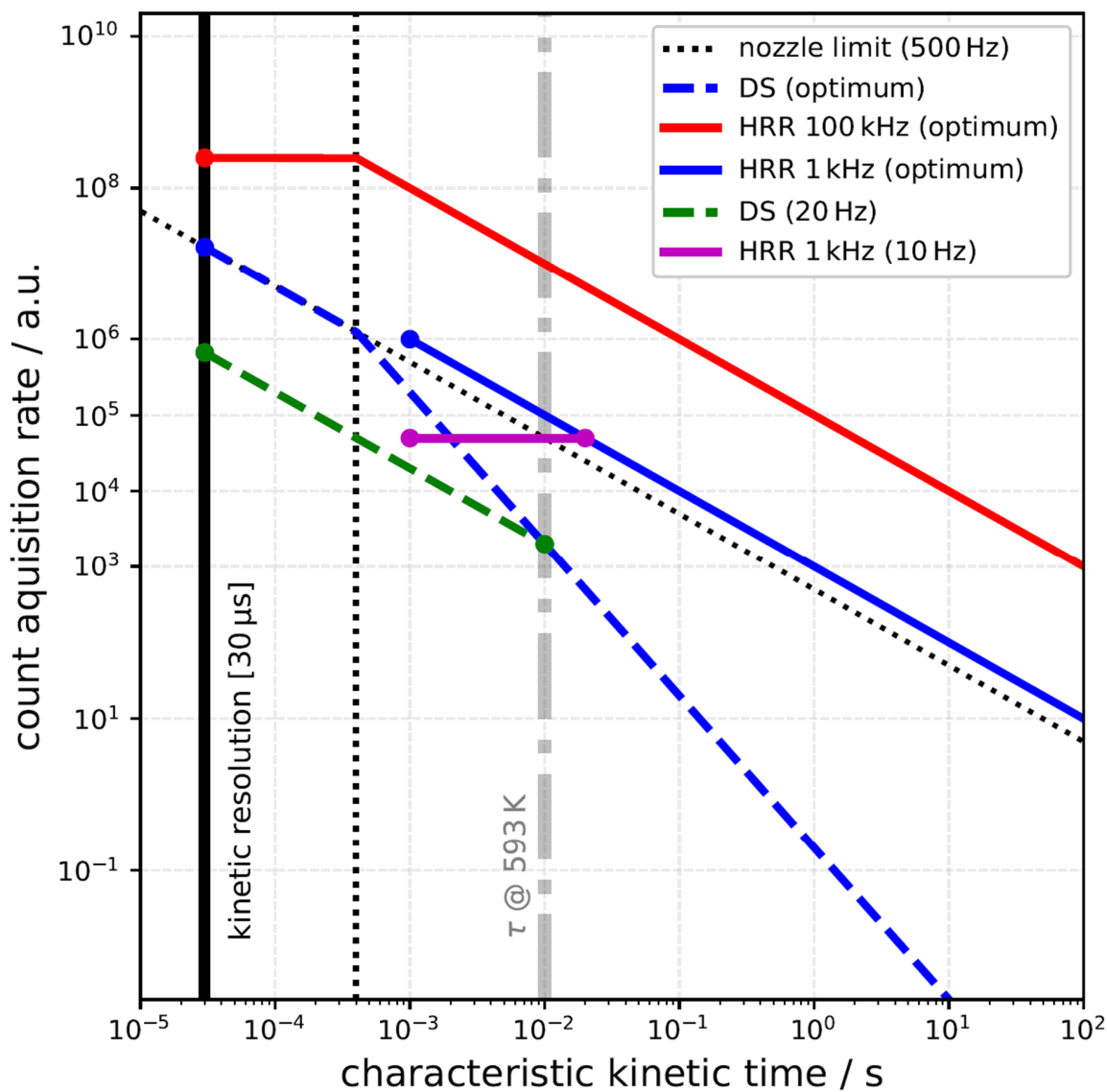

Fig. 4: Theoretical count acquisition rates (CAR) vs. characteristic kinetic time, $\tau$, for a variety of experimental configurations. Optimized delay-scanning (DS, blue dashed line) and 1−kHz high rep-rate detection (HRR, blue solid line) as well as an optimized high rep-rate detection with 100−kHz detection (red solid line) experiment are shown—here, only relevant data (see text) is obtained. Experimental configurations presented in this paper are also shown for delay-scanning (green dashed, CO nozzle at 20 Hz) and 1−kHz detection with CO beam operating at 10 Hz (magenta solid line). The gray dash-dotted line indicates the value of $\tau$ relevant to our experiments on CO trapping/desorption, where we measured the improvement to the CAR. The temporal resolution for a transient kinetics experiment is limited by the duration of the molecular beam pulse (black solid line). The minimum time between molecular beam pulses is limited by pumping speed and maximum pulsed valve frequency of 500 Hz (black dotted line).

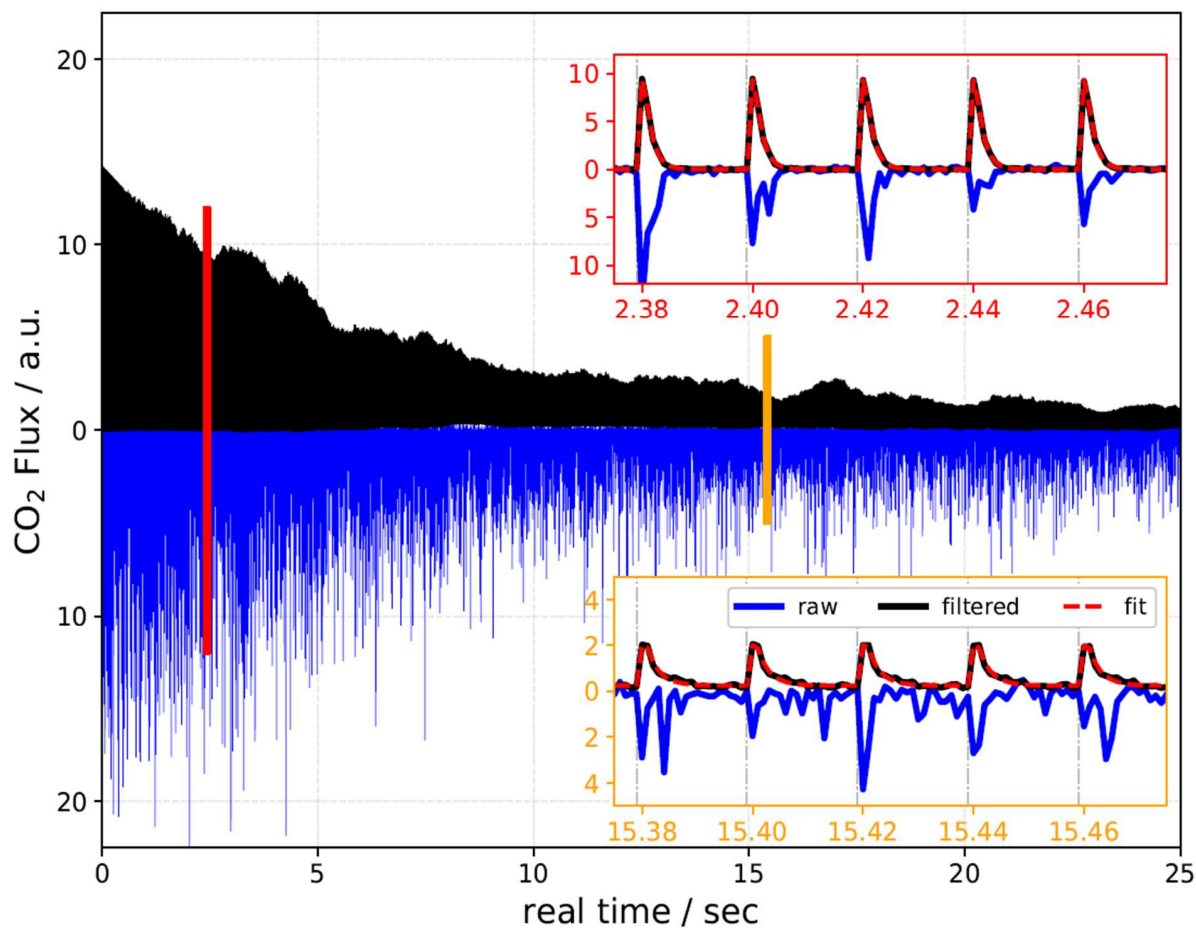

Fig 5: High rep-rate detection of velocity resolved kinetics for a non-stationery catalyst. The kinetics of CO oxidation on Pd(332) are recorded starting with saturated oxygen coverage. Adsorbed oxygen is removed during the experiment and the kinetics change accordingly. The surface temperature was 503 K and the CO beam operated at 50 Hz. The CO beam clean up a pre-oxidized surface that had been exposed to $300 \pm 80$ ML of $O_2$. The raw data are shown as blue lines, the Savitzky-Golay filtered data is shown as black lines. Kinetic fits (first order decay convoluted with incident beam shape) are shown as red dashed lines in the insets. The insets are indicated by colored bars and borders. The gray dash-dotted line in the insets indicates the reaction time at which the reaction is initiated by the pulsed CO beam.

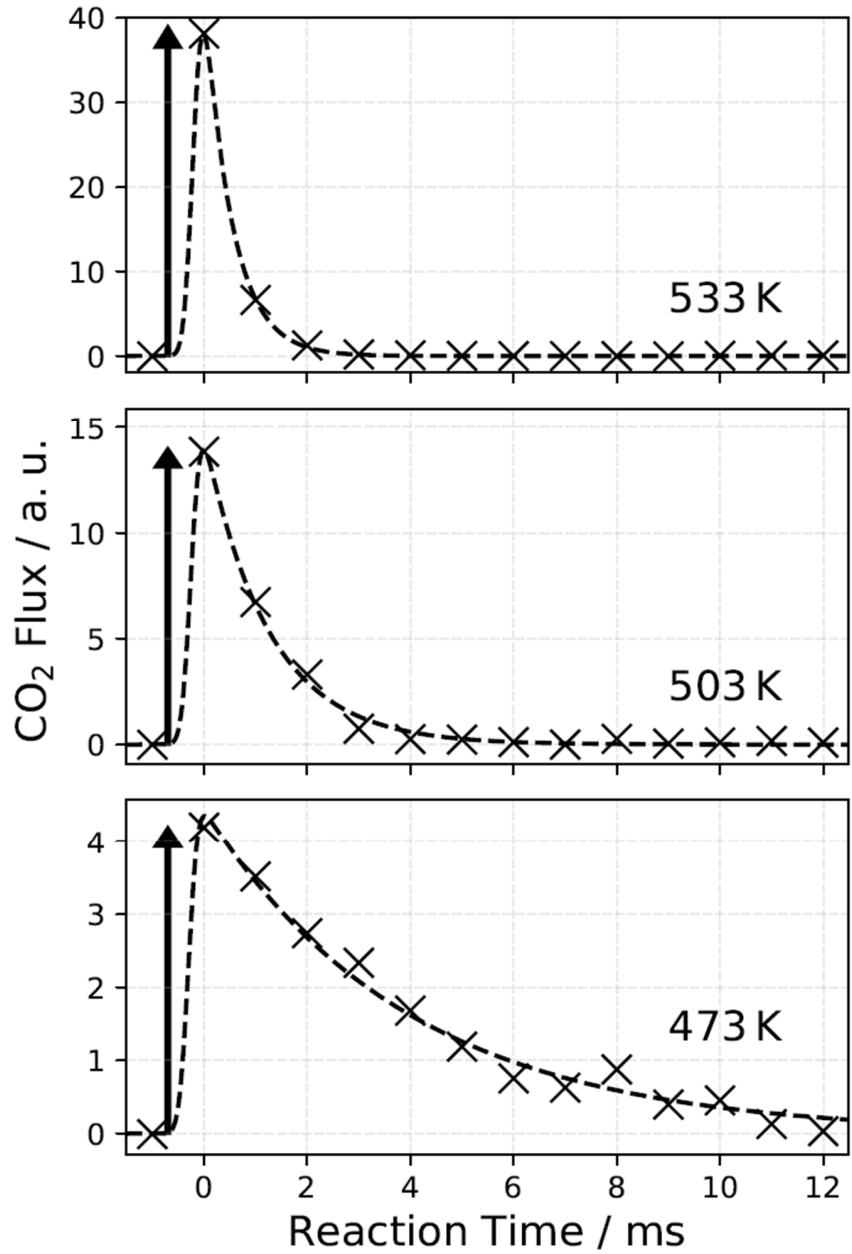

Fig 6: CO oxidation kinetics at saturated oxygen coverage. Kinetic traces (crosses) obtained by averaging over the first 20 pulses in experiments like those of Fig. 5. The dashed lines are fits to a first order decay (convoluted over the incident beam). The arrows indicate the relative initial rates in the three experiments.

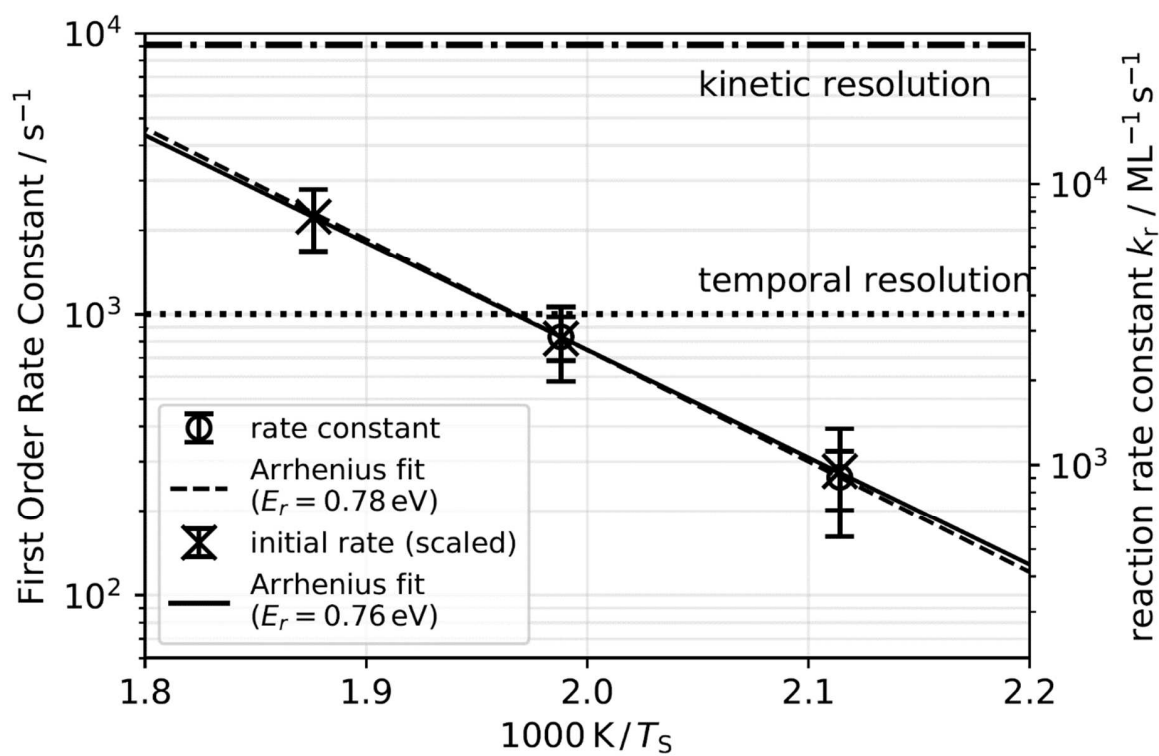

Fig 7: Temperature dependence of CO oxidation rate constants at saturated oxygen coverage. The first order rate constants for $CO_2$ formation determined from the data of Fig. 6. The circles are the rate constants determined from the shape of the single pulse kinetic trace and crosses are initial rates determined from their amplitude. The initial rates are scaled to match the first order rate constants at low temperature. The dotted line is the limit above which the rate constants cannot be derived from the shape of the kinetic trace. The dash-dotted line is the kinetic resolution (in this experiment around 110 µs) above which no kinetic information can be derived from transient kinetics. The full and dashed curves are Arrhenius fits to the initial rate and the first order rate constant, respectively.

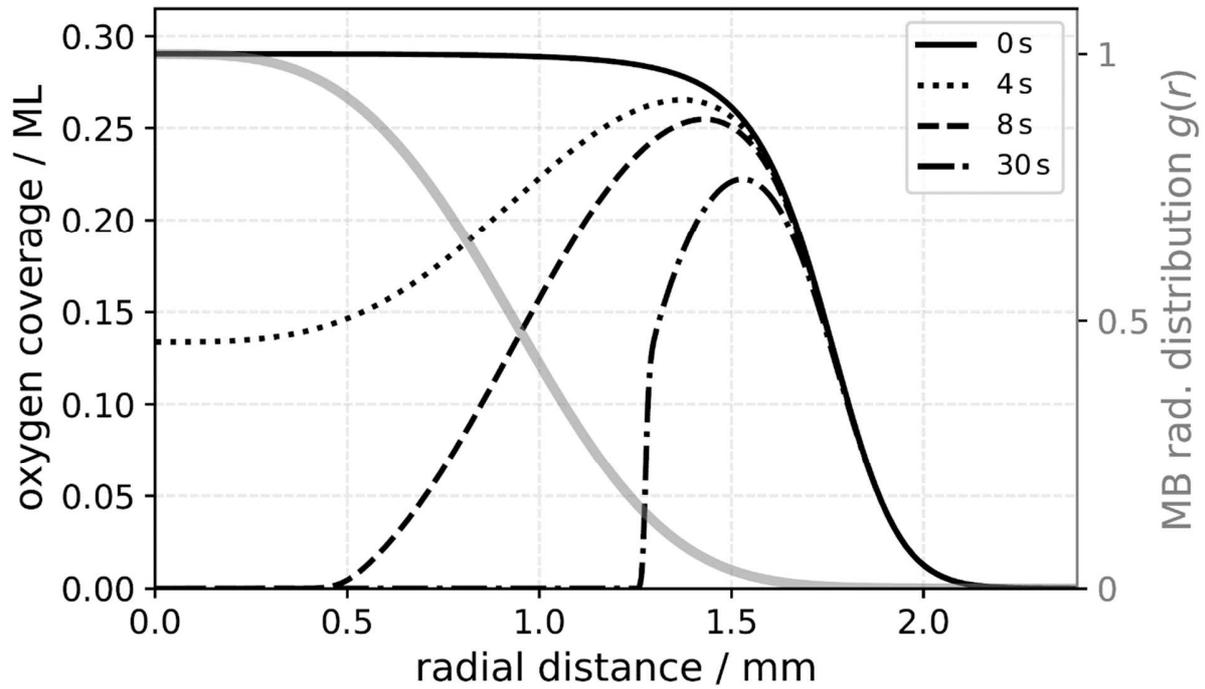

Fig. 8: Spatial distributions of adsorbed oxygen atoms during a CO oxidation titration. The distributions are assumed cylindrically symmetrical about the CO beam axis. The radial distance from the CO beam center-line is shown on the $x$-axis. The solid black line indicates the initial oxygen coverage distribution produced by long exposure with a molecular beam of $O_2$. The radial distribution of the CO beam (gray thick line) peaks near 0 and preferentially removes O-atoms there. As time progresses, a "doughnut hole" reaction develops, where the CO is concentrated along the CO beams center-line and adsorbed oxygen atoms form a ring around the CO beam. In later stages of titration, the reaction forms a front where the CO and O concentrations overlap. Diffusion of CO from the center of the doughnut hole to the oxygen ring also influences the reaction rate.

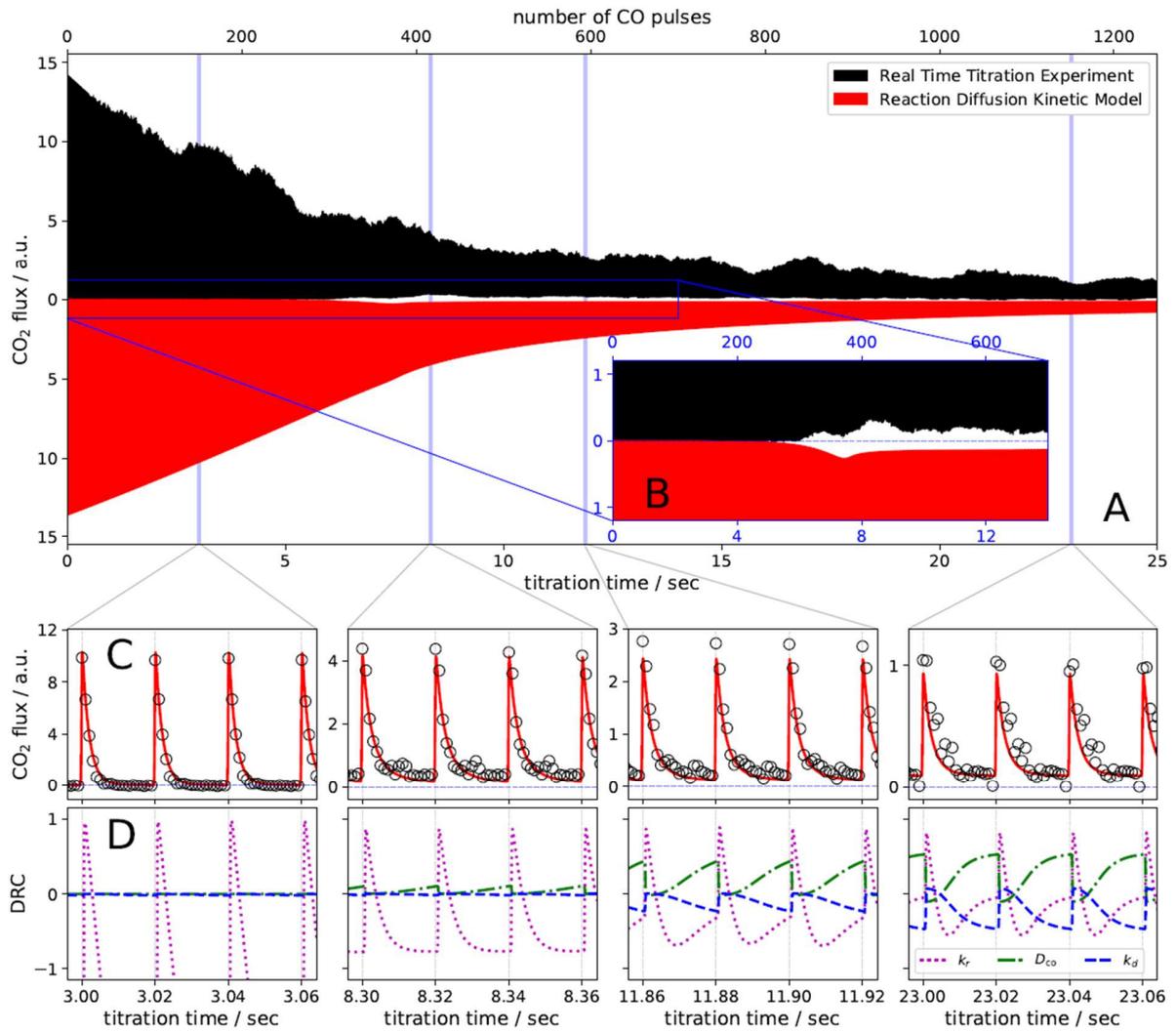

Fig. 9: Comparison of the model of real time titration (red solid lines) with measurements (black solid lines and open circles). The onset of the diffusion-controlled regime is indicated in panel B, where a continuous $CO_2$ production rate forms. Panels C show the results at 3.0, 8.3, 11.9 and 23.0 seconds after the start of the titration. The degree of rate control (DRC) is shown in panel D for three elementary processes: CO oxidation reaction (magenta dotted line), CO desorption (blue dashed line) and CO diffusion (green dash-dotted line). The DRC for O diffusion is for all conditions at least two orders of magnitude smaller and is therefore not shown. See text. Experimental conditions are as stated in Fig. 5.

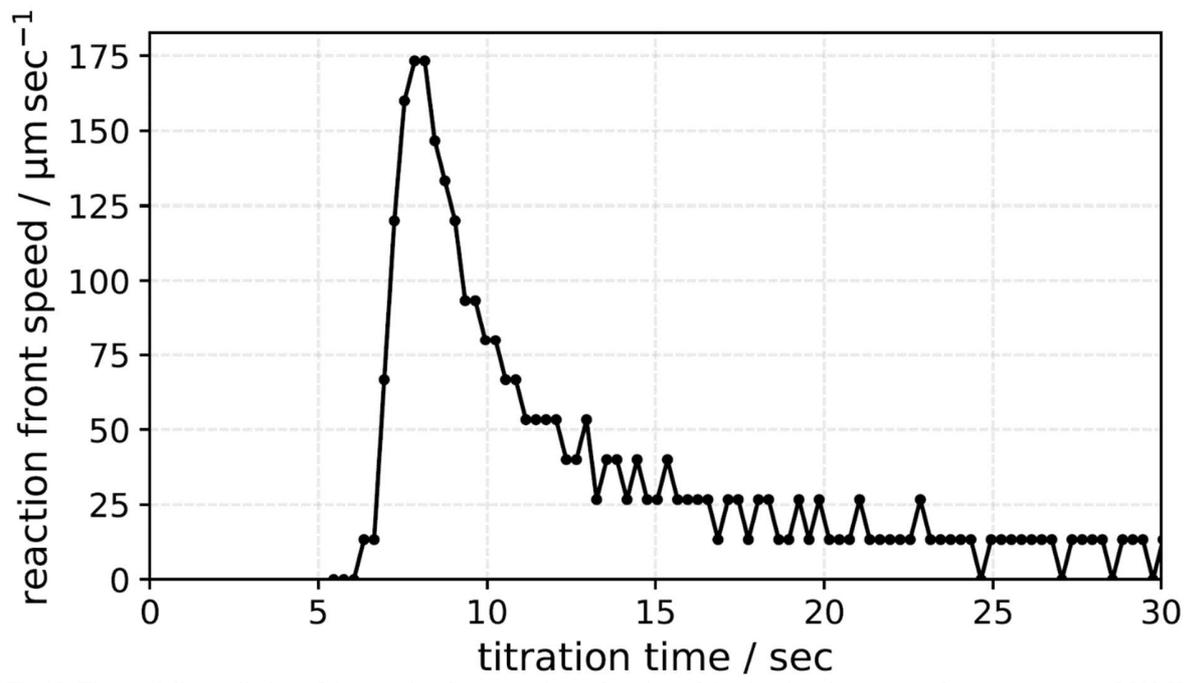
Fig 10: The model's prediction of the reaction front speed as a function of the titration time at a surface temperature of 503 K.

# 6. References

1. Schwarz, J. A.; Madix, R. J., Modulated beam relaxation spectrometry. *Surf Sci* **1974,** *46* (1), 317-341.

2. Gland, J. L.; Kollin, E. B., Carbon monoxide oxidation on the Pt(111) surface: Temperature programmed reaction of coadsorbed atomic oxygen and carbon monoxide. *The Journal of Chemical Physics* **1983,** *78* (2), 963-974.

3. Brown, L. S.; Sibener, S. J., A molecular beam scattering investigation of the oxidation of CO on Rh(111). I. Kinetics and mechanism. *The Journal of Chemical Physics* **1988,** *89* (2), 1163-1169.

4. Zaera, F., Use of molecular beams for kinetic measurements of chemical reactions on solid surfaces. *Surf Sci Rep* **2017,** *72* (2), 59-104.

5. Golibrzuch, K.; Shirhatti, P. R.; Geweke, J.; Werdecker, J.; Kandratsenka, A.; Auerbach, D. J.; Wodtke, A. M.; Bartels, C., CO desorption from a catalytic surface: elucidation of the role of steps by velocity-selected residence time measurements. *J Am Chem Soc* **2015,** *137* (4), 1465-75.

6. Neugebohren, J.; Borodin, D.; Hahn, H. W.; Altschaffel, J.; Kandratsenka, A.; Auerbach, D. J.; Campbell, C. T.; Schwarzer, D.; Harding, D. J.; Wodtke, A. M.; Kitsopoulos, T. N., Velocity-resolved kinetics of site-specific carbon monoxide oxidation on platinum surfaces. *Nature* **2018,** *558* (7709), 280-283.

7. Harding, D. J.; Neugebohren, J.; Hahn, H.; Auerbach, D. J.; Kitsopoulos, T. N.; Wodtke, A. M., Ion and velocity map imaging for surface dynamics and kinetics. *J Chem Phys* **2017,** *147* (1), 013939.

8. Harding, D. J.; Neugebohren, J.; Auerbach, D. J.; Kitsopoulos, T. N.; Wodtke, A. M., Using Ion Imaging to Measure Velocity Distributions in Surface Scattering Experiments. *J Phys Chem A* **2015,** *119* (50), 12255-62.

9. Park, G. B.; Kitsopoulos, T. N.; Borodin, D.; Golibrzuch, K.; Neugebohren, J.; Auerbach, D. J.; Campbell, C. T.; Wodtke, A. M., The kinetics of elementary thermal reactions in heterogeneous catalysis. *Nat Rev Chem* **2019,** *3* (12), 723-732.

10. Kondratenko, E. V.; Ovsitser, O.; Radnik, J.; Schneider, M.; Kraehnert, R.; Dingerdissen, U., Influence of reaction conditions on catalyst composition and selective/non-selective reaction pathways of the ODP reaction over V2O3, VO2 and V2O5 with O2 and N2O. *Applied Catalysis A: General* **2007,** *319*, 98-110.

11. Reuter, K.; Scheffler, M., Composition, structure, and stability ofRuO2(110)as a function of oxygen pressure. *Phys Rev B* **2001,** *65* (3).

12. Reuter, K.; Scheffler, M., First-principles atomistic thermodynamics for oxidation catalysis: surface phase diagrams and catalytically interesting regions. *Phys Rev Lett* **2003,** *90* (4), 046103.

13. Reuter, K.; Scheffler, M., Composition and structure of theRuO2(110)surface in anO2and CO environment: Implications for the catalytic formation ofCO2. *Phys Rev B* **2003,** *68* (4).

14. Beebe, T. P.; Goodman, D. W.; Kay, B. D.; Yates, J. T., Kinetics of the activated dissociative adsorption of methane on the low index planes of nickel single crystal surfaces. *The Journal of Chemical Physics* **1987,** *87* (4), 2305-2315.

15.     Savitzky, A.; Golay, M. J. E., Smoothing and Differentiation of Data by Simplified Least Squares Procedures. *Anal Chem* **1964,** *36* (8), 1627-1639.

16.     Engel, T., A molecular beam investigation of He, CO, and O2 scattering from Pd(111). *The Journal of Chemical Physics* **1978,** *69* (1).

17.     Conrad, H.; Ertl, G.; Koch, J.; Latta, E. E., Adsorption of CO on Pd single crystal surfaces. *Surf Sci* **1974,** *43* (2), 462-480.

18.     Engel, T.; Ertl, G., A molecular beam investigation of the catalytic oxidation of CO on Pd (111). *The Journal of Chemical Physics* **1978,** *69* (3), 1267-1281.

19.     Petzold, L., Automatic Selection of Methods for Solving Stiff and Nonstiff Systems of Ordinary Differential Equations. *Siam J Sci Stat Comp* **1983,** *4* (1), 136-148.

20.     Banse, B. A.; Koel, B. E., Interaction of oxygen with Pd(111): High effective O2 pressure conditions by using nitrogen dioxide. *Surf Sci* **1990,** *232* (3), 275-285.

21.     Guo, X.; Hoffman, A.; Yates, J. T., Adsorption kinetics and isotopic equilibration of oxygen adsorbed on the Pd(111) surface. *The Journal of Chemical Physics* **1989,** *90* (10), 5787-5792.

22.     Rose, M. K.; Borg, A.; Dunphy, J. C.; Mitsui, T.; Ogletree, D. F.; Salmeron, M., Chemisorption of atomic oxygen on Pd(111) studied by STM. *Surf Sci* **2004,** *561* (1), 69-78.

23.     Mitsui, T.; Rose, M. K.; Fomin, E.; Ogletree, D. F.; Salmeron, M., Diffusion and pair interactions of CO molecules on Pd(111). *Phys Rev Lett* **2005,** *94* (3), 036101.

24.     Campbell, C. T., The Degree of Rate Control: A Powerful Tool for Catalysis Research. *Acs Catal* **2017,** *7* (4), 2770-2779.

25.     Jakubith, S.; Rotermund, H. H.; Engel, W.; von Oertzen, A.; Ertl, G., Spatiotemporal concentration patterns in a surface reaction: Propagating and standing waves, rotating spirals, and turbulence. *Phys Rev Lett* **1990,** *65* (24), 3013-3016.

26.     Rotermund, H. H.; Jakubith, S.; von Oertzen, A.; Ertl, G., Solitons in a surface reaction. *Phys Rev Lett* **1991,** *66* (23), 3083-3086.

27.     Reutt-Robey, J. E.; Doren, D. J.; Chabal, Y. J.; Christman, S. B., CO diffusion on Pt(111) with time-resolved infrared-pulsed molecular beam methods: Critical tests and analysis. *The Journal of Chemical Physics* **1990,** *93* (12), 9113-9129.

28.     Blomberg, S.; Zetterberg, J.; Zhou, J.; Merte, L. R.; Gustafson, J.; Shipilin, M.; Trinchero, A.; Miccio, L. A.; Magaña, A.; Ilyn, M.; Schiller, F.; Ortega, J. E.; Bertram, F.; Grönbeck, H.; Lundgren, E., Strain Dependent Light-off Temperature in Catalysis Revealed by Planar Laser-Induced Fluorescence. *Acs Catal* **2016,** *7* (1), 110-114.

29.     Crank, J., *The Mathematics of Diffusion*. Clarendon Press: Oxford, 1975; Vol. 2.

30.     Serri, J. A.; Tully, J. C.; Cardillo, M. J., The influence of steps on the desorption kinetics of NO from Pt(111). *The Journal of Chemical Physics* **1983,** *79* (3), 1530-1540.

31. Šnábl, M.; Borusík, O.; Cháb, V.; Ondřejček, M.; Stenzel, W.; Conrad, H.; Bradshaw, A. M., Surface diffusion of CO molecules on Pd{111} studied with photoelectron emission microscopy. *Surf Sci* **1997,** *385* (2-3), L1016-L1022.

32. von Oertzen, A.; Rotermund, H. H.; Nettesheim, S., Investigation of diffusion of CO absorbed on Pd(111) by a combined PEEM/LITD technique. *Chem Phys Lett* **1992,** *199* (1-2), 131-137.

# 7. Appendix

## 7.1. Treatment of Diffusion

We use Fick's second law to account for CO diffusion which becomes important at later times in titration experiments. Assuming circular symmetry, Fick's law describes diffusion through a ring $r_j$ with Eq. I, written in polar coordinates.

$$\dot{n}_j = D_n \left( \frac{\partial^2 n}{\partial r^2}\bigg|_{r=r_j} + \frac{1}{r_j} \frac{\partial n}{\partial r}\bigg|_{r=r_j} \right) \tag{I}$$

In our application, the center of circular symmetry is crossing point of the center lines of the two molecular beams on the Pd surface. The coordinate system is sketched in Fig. A1. The CO beam axis is rotated by only 30° from the symmetry axis introducing a slightly elliptical CO spatial profile; but this is a small effect that we approximate as circular. We simulate the diffusion on the surface in a uniform sized radial grid, with $r_0$ being the radial width of each grid element and $r_j$ being the radial distance from center of circular symmetry (coordinate system origin) to the spatial element $j$.

$$r_j = \left(j + \frac{1}{2}\right) r_0 \tag{II}$$

Here, $j$ is an integer from 0, 1, …, $j_{\max}$.

Using finite differences, Eq. I can be expressed as:

$$\dot{n}_j = D_n \left( \frac{n_{j+1} + n_{j-1} - 2n_j}{r_0^2} + \frac{1}{r_j} \frac{n_{j+1} - n_{j-1}}{r_0} \right). \tag{III}$$

The result from Eq. III is identical to a formulation that can be found in Ref. [29], treating diffusion in a single component systems; however, we require a treatment of a two-component system. This is necessary because of site blocking effects that make the diffusion of two components dependent on one another. See Fig. A2.

Fick's law can be recast using the finite difference method (Eq. i**Error! Reference source not found.**) into a circularly symmetric grid-hopping formalism (comparable to Ref. [30]):

$$\dot{n}_j = k_{H,j+1} n_{j+1} + k_{H,j-1} n_{j-1} - k_{H,j} n_j - k_{H,j} n_j, \tag{IV}$$

where $k_{H,j}$ is the hopping rate constant of $n$ associated with the spatial element $j$.

We also need to consider that the perimeter of each grid element in polar coordinates increases with increasing radial distance—for example, the rate of hopping from $j$ to $j + 1$ will be favored over that from

$j$ to $j-1$ as the diffusion circumference is larger for the former. The resulting grid hopping rate constants are simply related to the diffusion coefficients:

$$k_{H,j+1} = \frac{D_n}{r_0^2}\left(1 + \frac{1}{2(j+0.5)}\right), \qquad (V)$$

$$k_{H,j} = \frac{D_n}{r_0^2}, \qquad (VI)$$

$$k_{H,j-1} = \frac{D_n}{r_0^2}\left(1 - \frac{1}{2(j+0.5)}\right). \qquad (VII)$$

The hopping transition rate for spatial element $j$ (Eq. IV) consists of 4 contributions: transitions from $j-1$ and $j+1$ to $j$ and trasnitions from $j$ to $j-1$ and $j+1$. The advantage formulating the diffusion problem in this way is that the diffusion of multiple species can be easily coupled to one another. This is done by introducing an occupation factor, $\gamma$, to each contribution of Eq. IV reflecting the binding site occupation of the spatial element to which the transition is described (e.g. to the term $k_{H,j-1}n_{j-1}$ the occupation of $j$ is relevant). The modified form of Eq. IV yields:

$$\dot{n}_j = k_{H,j+1}n_{j+1}\gamma_j^{(n)} + k_{H,j-1}n_{j-1}\gamma_j^{(n)} - k_{H,j}n_j\gamma_{j+1}^{(n)} - k_{H,j}n_j\gamma_{j-1}^{(n)}, \qquad (VIII)$$

where $\gamma_j^{(n)}$ is the occupation factor for the species $n$. We define $\gamma_j^{(n)}$ in the following way:

$$\gamma_j^{(n)} = \left(1 - \frac{m_j}{m_j^{max}}\right), \qquad (IX)$$

where $m_j$ is the concentration of species $m$ in spatial element $j$ and $m_j^{max}$ its concentration at maximum coverage. This approach means that the diffusive transport of species $n$ is hindered by coverages of species $m$. This is justified for CO and O on Pd, because both species are most stably bound at the same 3-fold site.

The rate formulation at the origin and outer edge of the coordinate system, i.e. at $j = 0$ and $j = j_{max}$, is given by Eq.'s X and XI.

$$\dot{n}_0 = 4k_{H,0}n_1\gamma_0^{(n)} - 4k_{H,0}n_0\gamma_1^{(n)} \qquad (X)$$

$$\dot{n}_{j_{max}}^{closed} = k_{H,j_{max}-1}n_{j_{max}-1}\gamma_{j_{max}}^{(n)} - k_{H,j_{max}}n_{j_{max}}\gamma_{j_{max}-1}^{(n)} \qquad (XI)$$

At $j_{max}$ (Eq. XI) the system may be closed—diffusion beyond this point is not possible—ensuring mass conservation as the system evolves. Alternatively, the system may be open—molecules can diffuse beyond $j_{max}$ but cannot return back, which yields Eq. XII:

$$\dot{n}_{j_{\max}}^{\text{open}} = k_{H,j_{\max}-1} n_{j_{\max}-1} \gamma_{j_{\max}}^{(n)} - k_{H,j_{\max}} n_{j_{\max}} - k_{H,j_{\max}} n_{j_{\max}} \gamma_{j_{\max}-1}^{(n)}. \qquad \text{(XII)}$$

We used the open termination condition in our analysis, but tested the closed termination. We found no differences because the simulated cell size was big enough.

### 7.2. Comparison of CO diffusion rates

In Fig. A3 we compare the fitted CO diffusion coefficient from this work to previous reports of CO diffusion rates on Pd(111). In previous work STM (Scanning Tunneling Microscopy) was used to measure the site-to-site hopping rate [23], whereas PEEM and LID [31-32] (Photo Electron Emission Microscopy and Laser induced desorption) provide a measure of macroscopic diffusion. Using our derived set of Arrhenius parameters, we reproduce previously reported diffusion constants well. This supports the idea that atomic steps of Pd(332) have no influence on the diffusive mobility of CO. This is consistent with our observations from desorption rate measurements that show no energetic preference for CO at steps of Pd(332). Although previous studies suggested that step influenced diffusion may be important for this system [31], we find no evidence for this. It is possible that use of a too tightly limited temperature range and/or approximations on the dimensionality of the diffusion problem lead to erroneous conclusions.

### 7.3. Estimation of the reaction front speed

In Fig. A4, we show the radial concentration profiles of CO* and O* at several late titration times. The rather steep decrease of both concentration profiles indicates a diffusion-controlled regime. The quasi-stationary $CO_2$ production is confined near the reaction front. In the titration experiment, the oxygen coverage steadily decays, while the CO coverage builds up. The reaction moves out from the center toward the outer flanks where the CO coverage steadily removes O*. An established characteristic for diffusion-controlled reactions is the reaction front propagation speed. We use plots like those shown in Fig. A4 to estimate the reaction front speed that is shown in Fig. 10.

## 8. Figures for Appendix

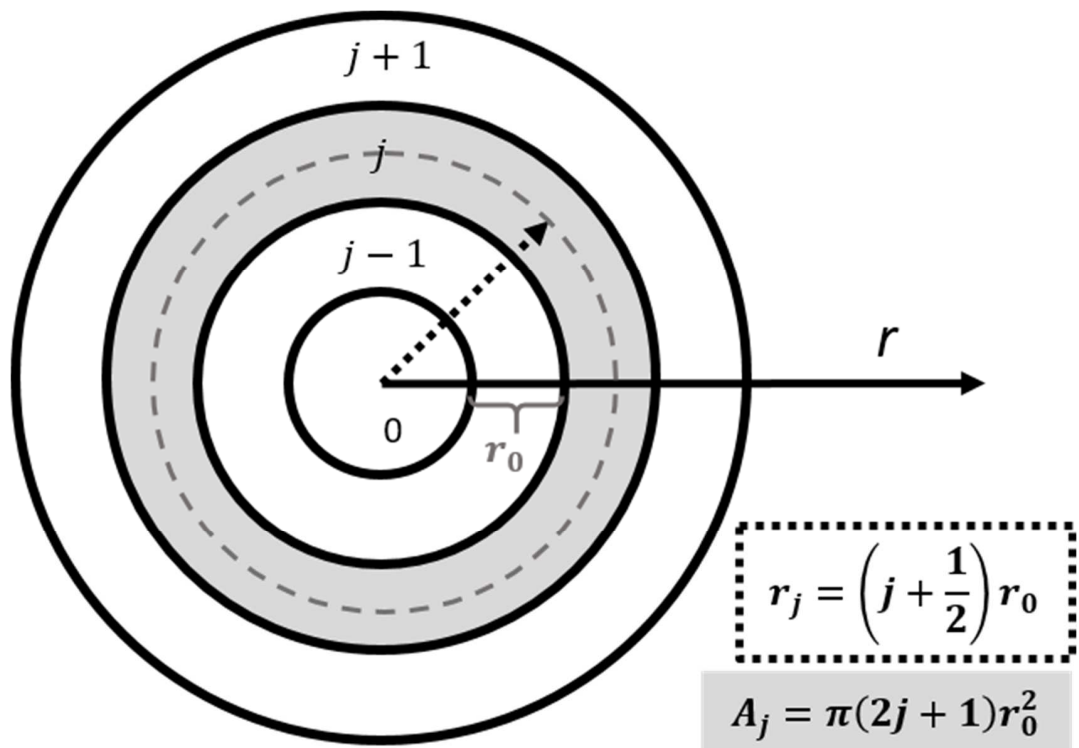

Fig A1: The coordinate system used for diffusion reaction modelling with key definitions.

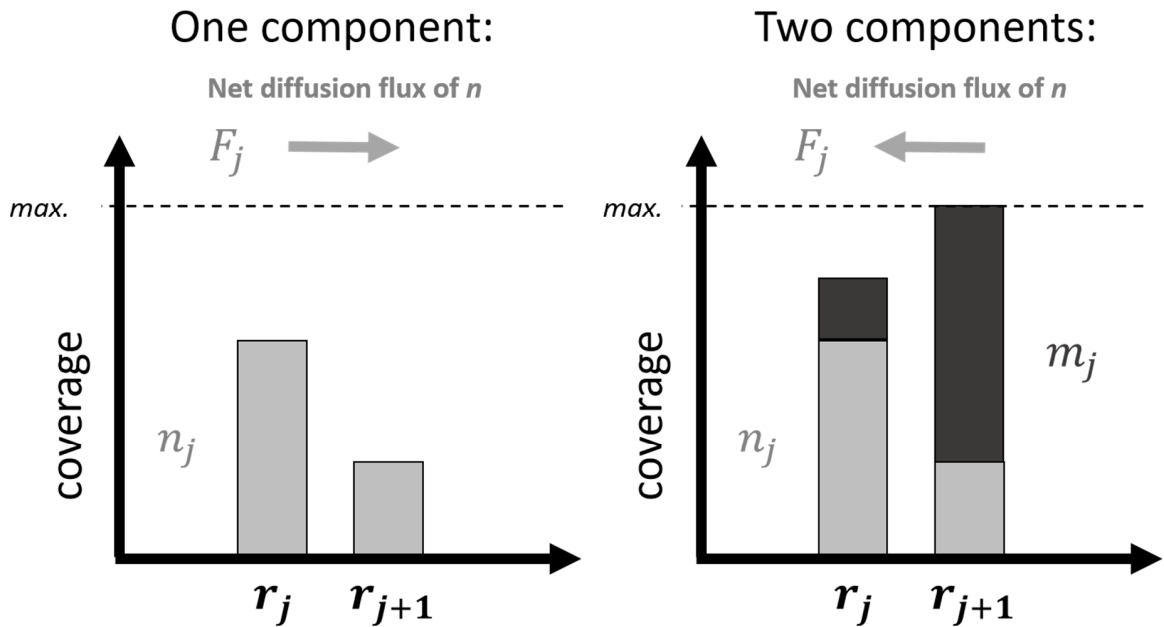

Fig A2: Illustration of the direction of the net diffusion flux of component *n* in a one-component system (left) and in a two-component system (right). In the one-component system the flux goes in the direction of steepest concentration gradient as expected from Fick's equations. However, when a second component *m* is present, this eventually leads to site blocking for component *n*. For situations as sketched on the right, only if the site blocking term $\gamma^{(n)}$ is introduced, which couples the concentrations of *n* and *m*, the right direction of the net diffusive flux is obtained for *n*.

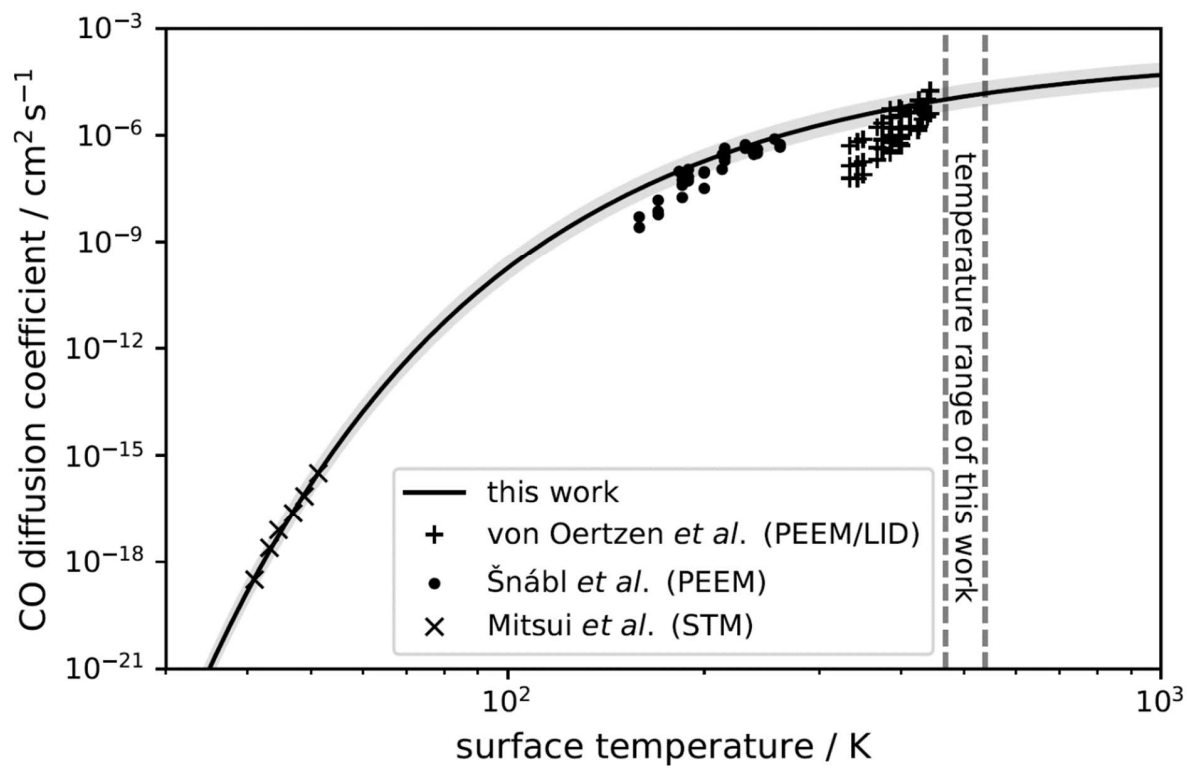

Fig A3: Thermal CO diffusion coefficients are shown for various studies (crosses, circles and pluses) on Pd(111) as a function of surface temperature. The methods by which the diffusion rates are determined are indicated in the legend. The results of this work are shown as solid black lines with grey shaded region indicating the estimated error range. The dashed vertical lines in the plot indicate the temperature range of this work.

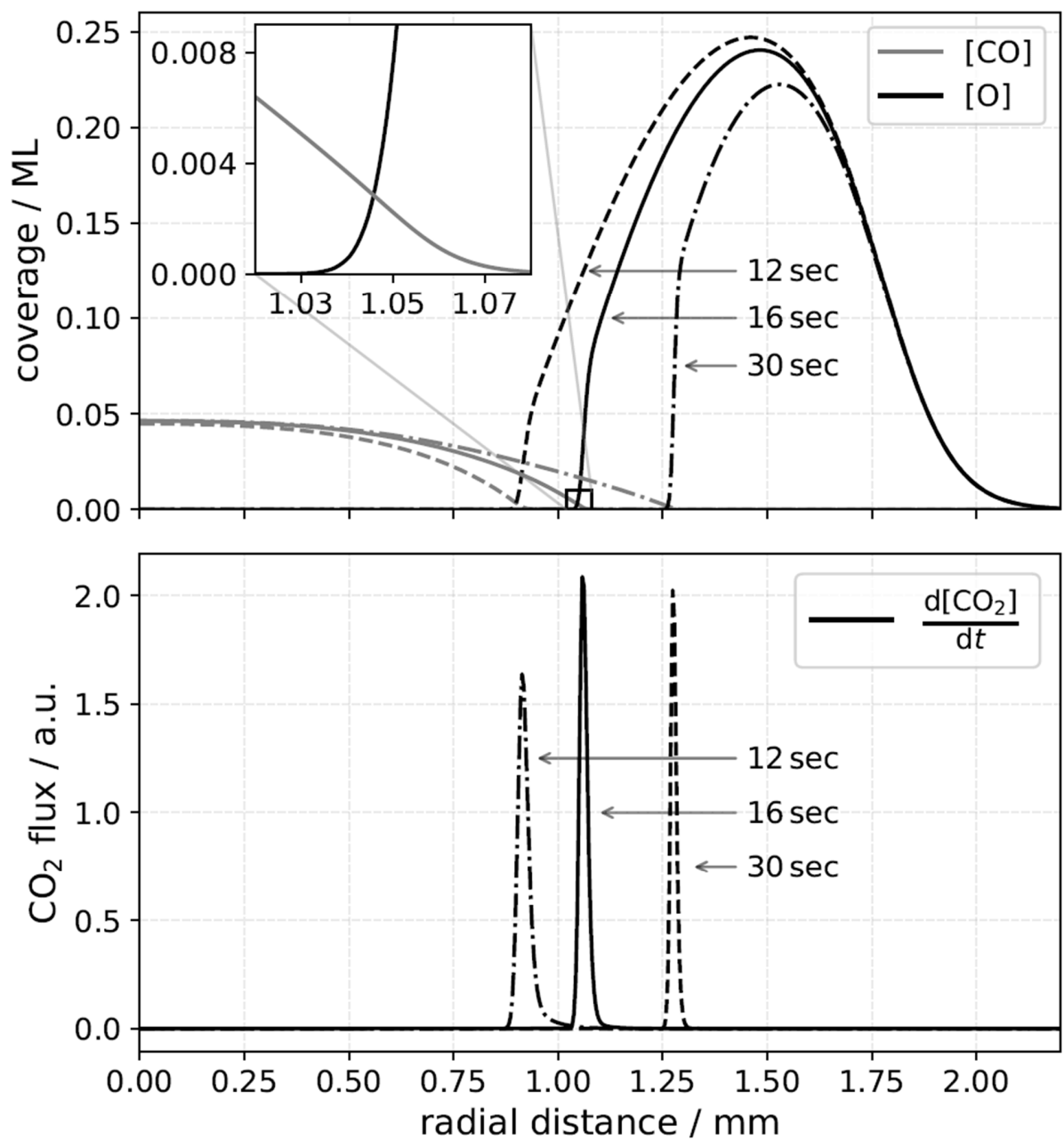

Fig. A4: Top: CO* and O* coverage profiles at late titration times at 503 K are shown. The inset of the top plot shows the intersection of the concentration profiles. Bottom: The stationary CO$_2$ formation rate as a function of the radial distance associated with the coverage profiles from the above plot. From the temporal evolution of the peak maximum the reaction front speed is estimated.